\documentclass[sigconf, screen]{acmart}
\AtBeginDocument{%
  }

\usepackage{array}
\usepackage{subfigure}
\usepackage{graphicx}
\usepackage[switch]{lineno}
\usepackage{amsmath}
\usepackage{amsthm}
\usepackage{multirow}
\usepackage{booktabs}
\usepackage{bbding}
\usepackage{algorithm}
\usepackage{algorithmic}
\usepackage{xcolor}
\usepackage{xspace}
\usepackage{arydshln}
\usepackage[inline]{enumitem}

\newcommand{\paratitle}[1]{\vspace{1.5ex}\noindent\textbf{#1}}
\newcommand{\ie}{\emph{i.e.,}\xspace}

\newcommand{\eg}{\emph{e.g.,}\xspace}

\newcommand{\fig}{Figure\xspace}

\usepackage[inline]{enumitem}

\makeatletter

\newcommand{\Rmnum}[1]{\expandafter\@slowromancap\romannumeral #1@}
\makeatother

\setcopyright{acmlicensed}
\copyrightyear{2026}
\acmYear{2026}
\acmDOI{XXXXXXX.XXXXXXX}
\acmConference[WSDM '26]{Proceedings of the 19th ACM International Conference on Web Search and Data Mining (WSDM '26)}{February 22-26, 2026}{Boise, Idaho, USA}
\acmISBN{978-1-4503-XXXX-X/2018/06}

\begin{document}

\title[Have We Really Understood Collaborative Information? An Empirical Investigation]{Have We Really Understood Collaborative Information? \\ An Empirical Investigation}


\author{Xiaokun Zhang}
\authornote{Xiaokun is also supported by the Key Laboratory of Social Computing and Cognitive Intelligence (Dalian University of Technology), Ministry of Education.}
\affiliation{%
  \institution{City University of Hong Kong}
  \city{Hong Kong}
  \country{China}
}
\email{dawnkun1993@gmail.com}

\author{Zhaochun Ren}
\authornote{Corresponding authors.}
\affiliation{%
  \institution{Leiden University}
  \city{Leiden}
  \country{The Netherlands}
  }
\email{z.ren@liacs.leidenuniv.nl}

\author{Bowei He}
\affiliation{%
  \institution{City University of Hong Kong}
  \city{Hong Kong}
  \country{China}
  }
\email{boweihe2-c@my.cityu.edu.hk}

\author{Ziqiang Cui}
\affiliation{%
  \institution{City University of Hong Kong}
  \city{Hong Kong}
  \country{China}
  }
\email{ziqiang.cui@my.cityu.edu.hk}

\author{Chen Ma}
\authornotemark[2]
\affiliation{%
  \institution{City University of Hong Kong}
  \city{Hong Kong}
  \country{China}
  }
\email{chenma@cityu.edu.hk}

\renewcommand{\shortauthors}{Xiaokun Zhang, et al.}

\begin{abstract}
Collaborative information serves as the cornerstone of recommender systems which typically focus on capturing it from user-item interactions to deliver personalized services. 
However, current understanding of this crucial resource remains limited. Specifically, a quantitative definition of collaborative information is missing, its manifestation within user-item interactions remains unclear, and its impact on recommendation performance is largely unknown.
To bridge this gap, this work conducts a systematic investigation of collaborative information. We begin by clarifying collaborative information in terms of item co-occurrence patterns, identifying its main characteristics, and presenting a quantitative definition. We then estimate the distribution of collaborative information from several aspects, shedding light on how collaborative information is structured in practice. 
Furthermore, we evaluate the impact of collaborative information on the performance of various recommendation algorithms. Finally, we highlight challenges in effectively capturing collaborative information and outlook promising directions for future research. 
By establishing an empirical framework, we uncover many insightful observations that advance our understanding of collaborative information and offer valuable guidelines for developing more effective recommender systems.
\end{abstract}

\ccsdesc[500]{Information systems~Recommender systems}

\keywords{Recommender systems, Collaborative information, Quantitative definition, Manifestation, Impact evaluation.}

\maketitle

\section{Introduction}

Recommender systems (RS) have become an indispensable tool in combating information overload in the current era of information explosion. They strive to infer users' interest from their history behaviors, and provide pertinent suggestions accordingly. Due to their remarkable potential in catering to individual preferences, RS have found applications across various domains, such as product recommendation in e-commerce~\cite{SASRec,FineRec}, music or video recommendation on multimedia platforms~\cite{Jiang@DASFAA2024, IP2}, and point-of-interest recommendation on location-based services~\cite{Li@SIGIR2024}.

The essence of RS is to uncover \textit{collaborative information} from massive user-item interactions to deliver personalized suggestions tailored to individual preferences~\cite{Shi@CSUR2014, Zhang@TKDE2025}. Current literature tends to interpret collaborative information as unique patterns embedded within user-item interactions~\cite{IKNN, Linden@InternetCompute2003}. A well-known example is the ``Beer and Diapers'', where these two seemingly unrelated items are often purchased together, prompting RS to recommend beer to users who have just bought diapers. Collaborative information has dominated the research of RS for decades. Over the years, a wide range of solutions has been proposed, spanning traditional techniques such as matrix factorization~\cite{SalakhutdinovNIPS2007} and nearest-neighbor-based methods~\cite{IKNN}, to modern neural network-based approaches like recurrent neural networks (RNN)~\cite{GRU4Rec, NARM}, attention mechanisms~\cite{SASRec, BERT4Rec} and graph neural networks (GNN)~\cite{SR-GNN, guo@WSDM2022}. More recently, substantial efforts are dedicated to integrating collaborative information into large language models (LLM) to unleash its potentials in recommendation tasks~\cite{TALLRec, Qu@RecSys2024, Zhang@SIGIR2025, Wang@WSDM2025}. 

Despite the significance of collaborative information in RS has been widely recognized~\cite{zhang2025shapley, David1992, Zhu@WWW2024, DIMO}, current understanding of this crucial resource remains largely vague and limited. In fact, a quantitative definition of collaborative information is absent, which hinders an in-depth exploration of this concept and how it manifests within user-item interactions. In addition, current methods primarily demonstrate their ability to capture collaborative information by reporting overall improvements in basic performance metrics, such as prediction precision. However, the specific impact of collaborative information on recommendation algorithms remains still unclear. This naturally leads to the central question posed in the title of this work: \textit{Have We Really Understood Collaborative Information?} Unfortunately, due to the limited research dedicated to this fundamental concept, the answer remains a resounding \textit{NO}.


To tackle this gap, this work systematically investigates collaborative information, seeking to address three key research questions:

\textbf{RQ-\Rmnum{1}}: \textit{How to define collaborative information quantitatively?}

RS typically aim at mining user or item similarity from user-item interactions to perform personalization. These similarities are often derived from item co-occurrence patterns, which serve as the guideline for their computation. For example, items that are frequently co-purchased by users are assumed to be similar, supporting the case of ``Beer and Diapers''. Therefore, we first clarify collaborative information in terms of item co-occurrence patterns. Afterwards, we identify its three key characteristics: Transitivity, Hierarchy, and Redundancy. With these principles in mind, we propose a quantitative definition for collaborative information, categorizing it into different types of collaborative relations.

\textbf{RQ-\Rmnum{2}}: \textit{How does collaborative information manifest within user-item interactions?}

To deepen the understanding of collaborative information, we explore the manifestation of collaborative relations within user-item interactions by examining their distributions from several aspects. 
To be specific, we begin by presenting the distribution of collaborative relations across all item pairs, highlighting the collaborative relations that RS would encounter during the training phase. Moreover, we analyze the distribution of collaborative relations relevant to label items, shedding light on the collaborative relations that RS need to predict during the inference phase. In addition, we discuss item co-occurrence frequencies within primary collaborative relations, revealing their long-tail features in distributions.


\textbf{RQ-\Rmnum{3}}: \textit{What is the impact of collaborative information on the performance of recommendation algorithms?}

To examine the capability of existing methods in capturing collaborative information, we evaluate a range of representative approaches, including traditional techniques, modern neural models and recent LLM-based solutions, across various settings. First, we present the overall performance of these approaches across various benchmarks, exploring the relationship between their accuracy and the distribution of collaborative relations. Second, we conduct experiments focusing on specific types of collaborative relations, highlighting performance differences of RS in handling diverse collaborative relations. Finally, we analyze the proportion of different collaborative relations in predictions generated by various methods, assessing RS ability to utilize diverse collaborative relations.

In addition, we also point out several challenges that remain unexplored in capturing collaborative information and outline promising avenues for effectively leveraging this crucial resource. In summary, our main contributions can be outlined as follows:


\begin{itemize}
    \item \textit{Definition and Evaluation}. We quantitatively define collaborative information, and evaluate its impact on the performance of recommender systems. To our best knowledge, this study marks the first attempt to systematically investigate this crucial resource in the community.
    \item \textit{Empirical study}. We conduct a series of experiments to explore the role of collaborative information in recommender systems. Our analysis uncovers many valuable insights, significantly enhancing our understanding of the collaborative information.
    \item \textit{Challenges and Future Directions}. We identify challenges to be solved in effectively handling collaborative information and point out several potential directions, offering valuable guidelines for future efforts in building more robust recommender systems.
\end{itemize}

\section{Definition of Collaborative Information (RQ-\Rmnum{1})}


\subsection{Problem Formulation}
This work adopts the setting of sequential recommendation to formulate the recommendation task. In this context, the goal is to predict a user's next interacted items based on those she has previously engaged with. Our reasons for the choice are mainly two-fold: (1) sequential recommendation has been a popular topic in RS and has dominated the community recently~\cite{Zhang@PR2026, BERT4Rec, Xia@WSDM2025}; and (2) this setting focuses on offering suggestions solely from user-item interactions represented by item IDs, contributing to exploring collaborative information.

Formally, $\mathcal{I}$ denotes the set of items, where $n = |\mathcal{I}|$ signifies the total number of items. A user-item interaction sequence $\mathcal{S}$ = [$x_1, x_2, \cdots, x_m$] is generated by a user over a certain period, with each $x_i$ $\in$ $\mathcal{I}$ and $m$ being the sequence length. All user-item interaction sequences form the sequence set $\mathcal{D}$, where $\mathcal{S} \in \mathcal{D}$. The goal of RS is to predict next interacted item $x_{m+1}$ based on the user's previous interaction sequence $\mathcal{S}$. 

\subsection{Clarification of Collaborative Information}

\begin{figure*}[t]
  \centering
  \includegraphics[width=0.95\linewidth]{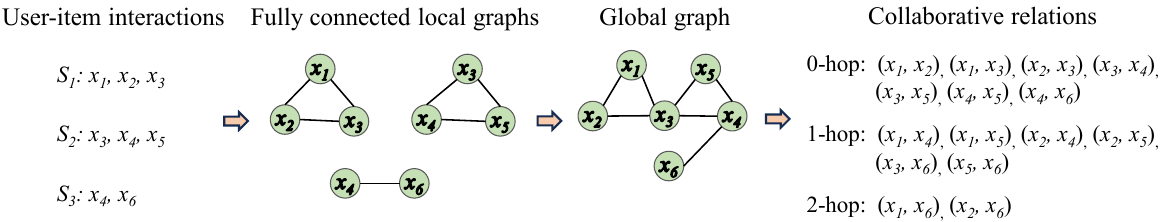}
  \caption{A quantitative definition of collaborative
information. Diverse collaborative relations are derived based on the complexity of item co-occurrence patterns.}\label{graphCI}
\end{figure*}

The concept of collaborative information can be traced back to Collaborative Filtering (CF) several decades ago~\cite{David1992, MDP}. CF-based methods can be classified into two types: user- or item-based CF. User-based CF finds similar users to make recommendations~\cite{Nathaniel@AAAI1999}, whereas item-based CF identifies similar items for this purpose~\cite{IKNN}. At their core, both approaches aim to mine similarity patterns embedded in item co-occurrence from user-item interactions~\cite{DIMO,Liu@CIKM2024}. Specifically, users' similarity is determined by the number of overlapped items in their interactions, while items' similarity is typically measured by their co-occurrence frequency across user-item interactions. 
In fact, RS methods, which primarily rely on user-item interaction data, could be viewed as leveraging item co-occurrence patterns to drive personalization. 

Consequently, this study concertizes collaborative information as item co-occurrence patterns. There are three primary benefits to this clarification of collaborative information: (1) it is intuitive and computationally efficient; (2) it offers a specific, quantitative representation of collaborative information; and (3) it aligns with the core principle of recommendation: leveraging collective insights from extensive user-item interactions to provide personalized suggestions for individuals. These benefits enable a finely-grained investigation of collaborative information, including its quantitative definition, manifestation analysis, and impact evaluation on RS.

\subsection{Characteristics of Collaborative Information}\label{chaofCR}
To intuitively illustrate item co-occurrence patterns, a toy example is plotted in \fig~\ref{graphCI}. In this example, each sequence is constructed as a fully connected local graph, capturing item co-occurrence relations within the sequence. These local graphs are then aggregated into a global graph, which represents co-occurrence relations from a global perspective. From the global graph, following characteristics of collaborative information can be observed:

\textbf{Transitivity} signifies that item similarity is transferable. For instance, $x_3$ has co-appeared with $x_1$ and $x_5$ in two unique sequences, implying that $x_1$ is similar to $x_3$, and $x_3$ is similar to $x_5$. Therefore, we can infer that $x_1$ is likely to have similar features to $x_5$.

\textbf{Hierarchy} reflects the layered structure of item co-occurrence relations. Taking the $x_1$ as an example, we can categorize other items into different groups based on the complexity of their co-occurrence relations: \{$x_2, x_3$\} directly connect with $x_1$ in the global graph; \{$x_4, x_5$\} connect with $x_1$ through one intermediate node; and \{$x_6$\} connects with $x_1$ through two intermediate nodes.

\textbf{Redundancy} refers to the presence of multiple co-occurrence relations between two items. For items $x_1$ and $x_2$, there are two paths connecting them: one direct path <$x_1-x_2$>, and another indirect one <$x_1-x_3-x_2$>.

\subsection{Quantitative Definition}\label{CRdefinition}

As previously discussed, the collaborative information can be concretized as item co-occurrence relations, which exhibit varying levels of complexity. To account for this, we categorize item collaborative information into different types based on the complexity, while highlighting its main characteristics. As illustrated in \fig~\ref{graphCI}, different types of collaborative relations (CR) are derived from the co-occurrence patterns observed in the global graph. Formally, these CRs can be formulated as follows:

\paratitle{0-hop CR}: if two items directly co-appear in a sequence, \ie neighbors in the global graph, they are defined to possess a 0-hop CR;

\paratitle{1-hop CR}: if two items don't co-appear in any sequences but are one-hop neighbors in the global graph, they are defined to possess a 1-hop CR;

\paratitle{2-hop CR}: if two items neither co-appear in any sequences nor share a one-hop CR but are two-hop neighbors in the global graph, they are defined to possess a 2-hop CR.

This formulation can be extended to define higher-order collaborative relations, such as 3-hop CR, 4-hop CR, and so on. The set of collaborative relations is defined as $\mathcal{R}$. Note that, our formal definition of CR aligns with the main characteristics of collaborative information. As shown in \fig~\ref{graphCI}, Transitivity is evident in 1-hop CR (\eg <$x_1-x_3-x_5$>), which are derived from two 0-hop CRs (<$x_1-x_3$> and <$x_3-x_5$>). The hierarchical nature of CR is also apparent, as these CRs are organized by their complexity levels. Additionally, we retain only one CR between any two items, prioritizing the nearest co-occurrence relation. For example, if there two paths between $x_1$ and $x_2$, namely <$x_1-x_2$> (0-hop CR) and <$x_1-x_3-x_2$> (1-hop CR), we retain the simpler direct CR. This approach ensures a concise yet meaningful representation of collaborative information. In this paradigm, we systematically capture and structure collaborative relations to serve as a foundation for further exploration.

Based on this quantitative definition of collaborative information, we can derive CRs between any two items within the item set $\mathcal{I}$. As illustrated in \fig~\ref{graphCI}, the process begins by constructing each user-item interaction sequence $\mathcal{S} \in \mathcal{D}$ as a fully connected local graph. These local graphs are then combined into a whole global graph by merging the nodes and edges of all the individual local graphs. Next, we apply a shortest path algorithm, like Dijkstra algorithm, to calculate the shortest path between every pair of items in $\mathcal{I}$. The length of the shortest path between two items serves as a proxy for the collaborative relation between them. For instance, if the shortest path length between two items is 1, it indicates a 0-hop CR between those items.

\begin{table*}[t]
\tabcolsep 0.22in 
\centering
\caption{Statistics of datasets.}
\begin{tabular}{c ccc ccc}
\toprule
Datasets      & Grocery & Cellphones & Cosmetics & Diginetica & Yoochoose & Tmall \\
\midrule
\#item        & 7,910   & 10,737   & 24,100  & 22,507   & 11,642    & 17,360\\
\#interaction & 152,102  & 150,568 & 406,214 &  349,959 & 868,812    & 172,028\\
\#sequence     & 40,534  & 46,315   & 45,893   & 67,350 & 194,992    & 21,811\\
avg.length    & 3.75     & 3.25     & 8.85    & 5.20    & 4.46    & 7.89\\
\bottomrule
\end{tabular}

\label{statistics}
\end{table*}

\section{Manifestation of Collaborative Information (RQ-\Rmnum{2})}

This section examines the distribution of collaborative relations within user-item interactions from various aspects, uncovering insights into the manifestation of collaborative information. We first introduce six benchmarks, all of which are commonly used to evaluate RS. The distribution of CR among all items is then presented. Following this, we analyze the CR distribution relevant to label items. Finally, we discuss item co-occurrence frequency within certain types of CRs. 

\subsection{Benchmarks}

In this study, following six popular datasets are employed to investigate the collaborative information:

\paratitle{Grocery} and \textbf{Cellphones} are two unique datasets sourced from Amazon\footnote{\url{http://jmcauley.ucsd.edu/data/amazon/}}. For a user, the user-item interaction sequence is constructed based on activities occurring within a single day, ensuring clear item co-occurrence patterns~\cite{MMSBR,DIMO}. 

\paratitle{Cosmetics}\footnote{\url{https://www.kaggle.com/mkechinov/ecommerce-events-history-in-cosmetics-shop}} contains five months of user behavior data from a medium-sized cosmetics online store. Following~\cite{Chen@SIGIR2023, CoHHN, BiPNet}, interactions from October 2019 labeled with the `add\_to\_cart' or `purchase' behavior types are extracted as the dataset.

\paratitle{Diginetica}\footnote{\url{https://competitions.codalab.org/competitions/11161}}, originally parts of the CIKM Cup 2016 competition, comprises user-item interactions collected from e-commerce search engine logs~\cite{DHCN}. 

\paratitle{Yoochoose}\footnote{\url{http://2015.recsyschallenge.com/challege.html}} includes six months of click-stream data from a European e-commerce website. Similar to prior research~\cite{GRU4Rec,Chang@WSDM2024}, we use a 1/64 subsample of this dataset to formulate user behaviors.


\paratitle{Tmall}\footnote{\url{https://tianchi.aliyun.com/dataset/dataDetail?dataId=42}} consists of user shopping logs from the Chinese e-commerce platform--Tmall. A user's `click' behaviors within one day are converted into a user-item interaction sequence~\cite{Han@SIGIR2022,Qiao@CIKM2023}.

For all benchmarks, following standard preprocessing steps~\cite{NARM,BERT4Rec,guo@WSDM2022}, sequences of length 1 and items with fewer than five occurrences are excluded. In user-item interaction sequences, the last item in each sequence serves as the prediction label, and the remaining items are used to capture user preferences. For example, a sequence [$x_1, x_2, ..., x_t$] is split as ([$x_1, x_2, ..., x_{t-1}$], $x_t$). Each dataset is chronologically split into three parts for training, validation, and test sets in a 7:2:1 ratio. We also filter out items that do not appear in the training set from validation and test sets. Table~\ref{statistics} summarizes the statistical details of these datasets.

\subsection{CR Distribution among Items}

In the current practice of the recommendation task, the training data is used to update model parameters during the training phase, while the test data is employed to evaluate model performance in the inference stage. In such a paradigm, RS methods rely on the training data to learn recommendation knowledge, \eg collaborative information. Therefore, we analyze the distribution of CR within the training data, shedding light on the CRs that RS would encounter during the learning phase. 

Following the definition of CR provided in Section~\ref{CRdefinition}, we examine the CR between all pairs of items in the item set, using the user-item interaction sequences from the training set. Afterwards, we count the frequency of each type of CR occurring in all item pairs, generating CR distribution among items. This distribution across six benchmarks is presented in~\fig~\ref{overallD}, where following observations are noted:

\begin{figure}[t]
  \centering
  \includegraphics[width=0.99\linewidth]{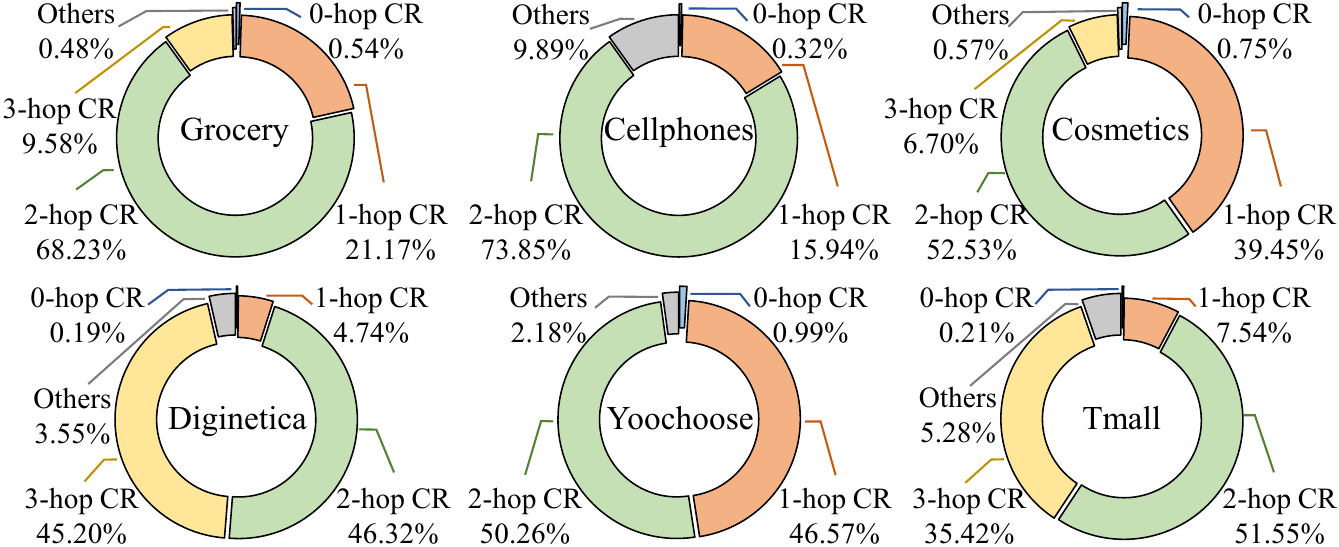}
  \caption{Distribution of collaborative relations among items.}\label{overallD}
\end{figure}

First, the direct CR, \ie 0-hop CR, accounts for only a small proportion of all item-item CRs, typically less than one percent. This indicates that \textbf{only a few pairs of items are usually co-purchased by users.} The scarcity of direct CR between items makes it particularly valuable for generating recommendations, leading to the effectiveness of cases like ``Beer and Diapers''.

Second, indirect CRs dominate CRs between items. As shown in the figure, 1-hop and 2-hop CRs together constitute around 90\% of all collaborative relations in most benchmarks. This highlights that \textbf{RS primarily encounters indirect CRs in real-world scenarios}, requiring them to learn recommendation knowledge about these complex relations to deliver personalized results.

Third, there is \textbf{a positive correlation between the prevalence of adjacent levels of CRs.} For example, benchmarks with a higher volume of 0-hop CR tend to contain much more 1-hop CR. This phenomenon aligns with the transitivity characteristic of collaborative information, where simpler CRs give rise to more complex ones. This alignment supports the effectiveness of our definition for collaborative information as presented in Section~\ref{CRdefinition}.

Finally, \textbf{the distribution of CRs varies across diverse benchmarks.} Taking the 2-hop CR as an example, it dominates other types of CRs in most datasets. However, the proportions of 2-hop CR and 3-hop CR are comparable in Diginetica. Moreover, the proportion of 1-hop CR differs significantly across datasets, such as 46.57\% in Yoochoose versus 4.74\% in Diginetica. These variations in CR distributions can be attributed to dataset-specific features, such as item volume, sequence length, and other factors, as shown in Table~\ref{statistics}. Additionally, these differences in CR distributions across benchmarks lead to the varying performance of RS models, which will be discussed in detail in Section~\ref{impactofcr}.

\subsection{CR Distribution relevant to Label Items}\label{CRdistribution}

After analyzing the CR distribution which RS encounter during the training phase, we explore the CRs that RS need to predict in the inference stage. Specifically, we begin by computing the collaborative relations of all item pairs based on user-item interaction sequences from the training set, generating the CR record. Next, for each test sample ([$x_1, x_2, \cdots, x_{t-1}$], $x_t$) in the test set, we retrieve the corresponding CR between the label item $x_t$ and each item in the sequence using the CR record. This results in a label CR sequence [$r_{t,1},r_{t,2},\cdots,r_{t,t-1}$] where $r_{i,j} \in \mathcal{R}$. Finally, we count the frequency of each type CR appearing across all test samples (\ie all label CR sequences in the test set). The CR distribution relevant to label items in various benchmarks is shown in~\fig~\ref{labelD}, where we can make the following observations:

\begin{figure}[t]
  \centering
  \includegraphics[width=0.98\linewidth]{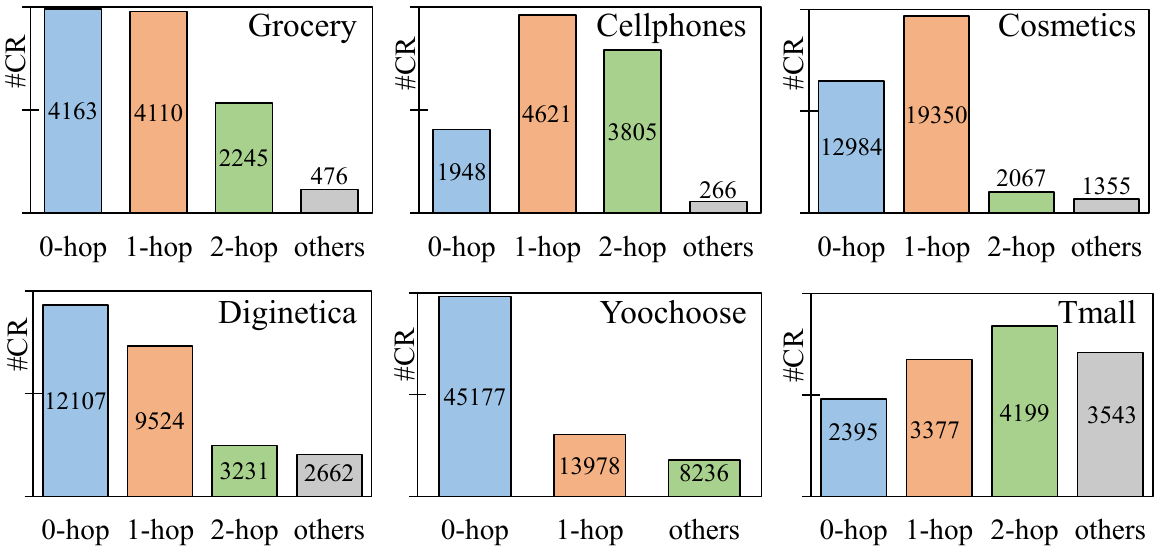}
  \caption{Collaborative relation distribution relevant to label items within various benchmarks.}\label{labelD}
\end{figure}

First, simple CRs (such as 0-hop and 1-hop CRs) between label items and items in the corresponding sequences are generally much more prevalent than complex ones. This indicates that \textbf{a model's ability to effectively handle simple CRs is of importance to its overall recommendation performance.} Since simple CRs are relatively easy to capture, basic methods can detect and leverage them for making recommendations. This supports the phenomenon that traditional techniques, like heuristic methods based on nearest-neighbor, can sometimes achieve performance comparable to more advanced models, such as neural networks~\cite{Garg@SIGIR2019, Maurizio@RecSys2019}.

Second, 0-hop CR, \ie direct CR, occurs frequently in user-item interactions, even though it constitutes a small proportion of all item-item CRs as shown in~\fig~\ref{overallD}. This suggests that \textbf{the direct co-occurrence relation frequently occurs within a very limited set of item pairs}, making it highly representative and significant. This finding further reinforces the rationality behind the classic ``Beer and Diapers'' example.

Finally, there are very few cases, none in most benchmarks, where the label item lacks any CRs with items in the corresponding sequence. As a result, we omit this situation in~\fig~\ref{labelD}. This observation indicates that the CRs identified in training data cover the situation in test data. In other words, \textbf{the collaborative information extensively exists in user-item interactions, presenting consistency between training and test data.} This consistency is the key reason why RS can work, where models trained on the training data can generalize effectively to make personalized recommendations on the test data.

\subsection{Item co-occurrence frequency in CRs}

\begin{figure}[t]
  \centering
  \includegraphics[width=0.98\linewidth]{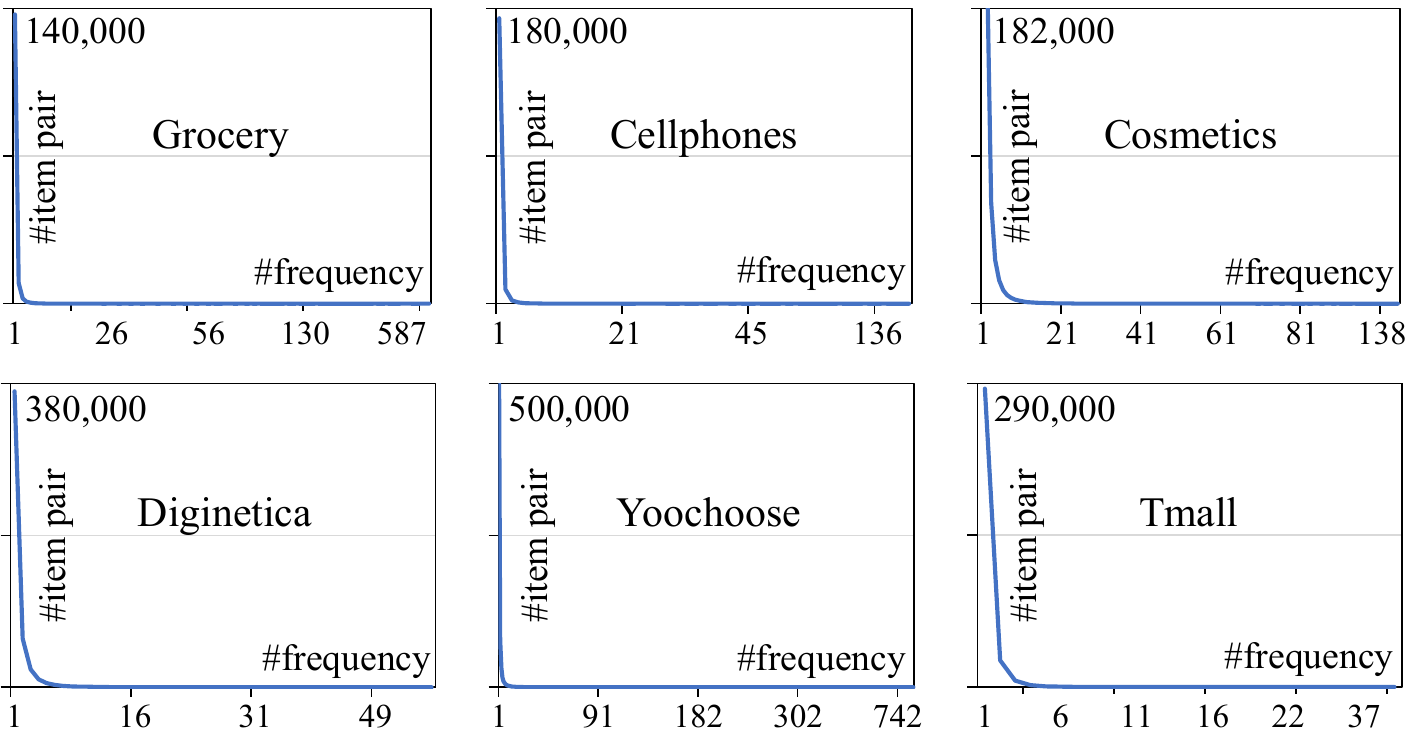}
  \caption{Item co-occurrence frequency within 0-hop CR on various benchmarks.}\label{co_0}
\end{figure}

In this section, we further delve into certain types of CRs, investigating item co-occurrence frequency. Since the 0-hop CR is easy to interpret and serves as the basis for all other types of CR, we focus our analysis on 0-hop CR. Specifically, we count the co-occurrence frequency of each item pair with a 0-hop CR in the training data. The results are presented in~\fig~\ref{co_0}, where the X-axis represents the co-occurrence frequency of item pairs with 0-hop CR, and the Y-axis shows the number of corresponding item pairs. 

From~\fig~\ref{co_0}, we observe that the item co-occurrence frequency in 0-hop CR follows a long-tail distribution. This indicates that \textbf{most items just co-occur in a few of sequences}, meaning that cases like ``Beer and Diapers'' are exceptions. This distribution highlights the inherent sparsity and imbalance in item co-occurrence relations, underscoring the challenges involved in effectively capturing collaborative information.

\section{Impact of Collaborative Information on RS (RQ-\Rmnum{3})}\label{impactofcr}
This section aims to explore the impact of collaborative information on RS performance. Specifically, we evaluate a range of representative RS methods across different settings of collaborative relations. At first, we describe the experiment setup in this work, including the studied methods, evaluation ‌protocol and implementation details. Based on the overall performance of studied methods, we discuss how the distribution of CRs influences RS performance. Subsequently, we examine these methods' performance on different types of CRs, identifying performance variations of RS in handling different CRs. Especially, we delve into the situation on direct CR versus in-direct CRs. Finally, we explore the proportion of various CRs within the predictions produced by RS methods, assessing RS ability to leverage collaborative relations.

\begin{table*}[ht]
\tabcolsep 0.03in 
  \centering
    \caption{Overall Performance of all methods on six benchmarks. The superscript in each model's result(\%) represents its ranking within the respective column, with higher rankings indicating better performance.}
    \begin{tabular}{c cccc cccc cccc}  
    \toprule  
    \multirow{2}*{Method}& 
    \multicolumn{2}{c}{Grocery}&\multicolumn{2}{c}{Cellphones}&\multicolumn{2}{c}{Cosmetics}&\multicolumn{2}{c}{Diginetica}&\multicolumn{2}{c}{Yoochoose}&\multicolumn{2}{c}{Tmall}\cr  
    \cmidrule(lr){2-3} \cmidrule(lr){4-5} \cmidrule(lr){6-7} \cmidrule(lr){8-9} \cmidrule(lr){10-11} \cmidrule(lr){12-13}
    &Prec@10&MRR@10 &Prec@10&MRR@10 &Prec@10&MRR@10 &Prec@10&MRR@10 &Prec@10&MRR@10 &Prec@10&MRR@10\cr  
    \midrule  
    BPR-MF      &$12.60^{12}$&$3.36^{12}$   &$1.04^{12}$&$0.21^{12}$    &$3.37^{11}$&$1.21^{12}$    &$2.65^{12}$&$0.94^{12}$    &$13.36^{12}$&$5.03^{12}$   &$6.23^{12}$&$2.33^{12}$\cr 
    Item-KNN    &$29.62^{11}$&$19.50^{10}$  &$8.04^{11}$&$5.70^{11}$    &$16.38^7$&$7.14^9 $        &$32.20^6$&$12.35^7 $       &$45.99^{10}$&$21.74^{10}$  &$20.67^9$&$10.43^{10}$\cr 
    SKNN        &$39.21^5$&$16.33^{11}$     &$14.64^7$&$6.28^{10}$      &$19.90^5$&$7.88^7 $        &$33.95^5$&$13.06^4 $       &$50.35^8$&$23.21^9$        &$23.37^8$&$14.87^{7}$\cr  
    GRU4Rec     &$33.61^{10}$&$29.46^9$     &$8.95^{10}$&$6.79^9$       &$10.65^{10}$&$7.53^8 $     &$17.50^{11}$&$8.27^{11}$   &$38.35^{11}$&$20.66^{11}$  &$12.30^{11}$&$8.36^{11}$\cr
    NARM        &$35.38^9$&$31.06^6$        &$15.90^6$&$13.88^1$        &$12.62^9$&$6.90^{10} $     &$23.05^9$&$10.85^9$        &$55.08^6$&$28.97^5$        &$24.99^6$&$19.42^4$\cr
    BERT4Rec    &$43.72^2$&$36.05^2$        &$17.76^3$&$10.84^5$        &$29.73^3$&$14.62^3$        &$34.05^4$&$12.99^5  $      &$59.85^3$&$30.11^3$        &$24.00^7$&$12.79^8$\cr  
    SR-GNN      &$35.81^8$&$30.56^7$        &$14.34^8$&$10.02^7$        &$2.87^{12}$&$1.24^{11}$    &$19.77^{10}$&$9.60^{10}$   &$54.58^7$&$27.97^7$        &$17.97^{10}$&$10.47^9$\cr
    LESSR       &$43.28^3$&$35.67^3$        &$16.88^4$&$10.95^4$        &$32.05^2$&$17.26^2$        &$36.67^2$&$14.75^2$        &$61.39^2$&$32.25^2$        &$28.70^5$&$18.56^5$\cr
    DHCN        &$40.98^4$&$30.29^8$        &$18.53^2$&$10.32^6$        &$25.43^4$&$12.26^4$        &$35.74^3$&$14.25^3$        &$57.18^4$&$28.15^6$        &$37.14^3$&$22.52^3$\cr
    MSGIFSR     &$44.77^1$&$36.13^1$        &$20.64^1$&$12.98^2$        &$34.77^1$&$18.15^1$        &$41.70^1$&$16.89^1$        &$62.68^1$&$33.30^1$        &$39.83^1$&$24.90^1$\cr
    Atten-Mixer &$38.67^6$&$32.30^4$        &$16.53^5$&$11.52^3$        &$17.62^6$&$9.32^6$         &$28.16^7$&$11.79^8$        &$56.02^5$&$29.63^4$        &$29.45^4$&$17.90^6$\cr
    P5          &$38.43^7$&$32.15^5$        &$11.16^9$&$9.19^8$         &$16.27^8$&$10.12^5$        &$26.56^8$&$12.70^6$        &$46.66^9$&$27.41^8$        &$38.62^2$&$24.81^2$ \cr
    
    
    \bottomrule
    \end{tabular}
    \label{performance}
\end{table*}

\subsection{Experiment Setup}

\subsubsection{Studied Methods}
We select a range of representative recommendation methods to examine the impact of CRs on recommendation performance, including traditional techniques, modern neural models, and recent LLM-based solutions. 
Note that, the primary goal of this work is to investigate the role of collaborative information embedded in user-item interactions in making recommendations. Therefore, we focus on methods that generate personalized recommendations solely based on user-item interactions (\ie item ID sequences), excluding the influence of other semantic information (\eg item text or images) on model predictions. 

\paratitle{Traditional techniques:} (1) \textbf{BPR-MF}~\cite{BPR-MF} leverages Matrix Factorization to mine collaborative information from user-item interaction matrix; (2) \textbf{Item-KNN}~\cite{IKNN} computes the similarity of items based on their co-occurrence frequency and generates recommendations accordingly; (3) \textbf{SKNN}~\cite{Jannach@RecSys2017} determines next items based on sequence (user) similarity.

\paratitle{Neural networks:} (4) \textbf{GRU4Rec}~\cite{GRU4Rec} leverages Gated Recurrent Units (GRU) to mine sequential transition patterns between items; (5) \textbf{NARM}~\cite{NARM} employs GRU with the attention mechanism to capture a user’s main intention; (6) \textbf{BERT4Rec}~\cite{BERT4Rec} relys on a bidirectional self-attention architecture to encode users' history behaviors; (7) \textbf{SR-GNN}~\cite{SR-GNN} converts user-item interactions into a graph and utilizes GNN to capture the pairwise transitions between items; (8) \textbf{LESSR}~\cite{LESSR} handles information loss of GNN-based models by introducing shortcut path in a graph and highlighting temporal order of items in information aggregation; (9) \textbf{DHCN}~\cite{DHCN} captures beyond-pairwise relations among items by hypergraph; (10) \textbf{MSGIFSR}~\cite{guo@WSDM2022} divides a sequence into multiple snippets to capture fine-grained co-occurrence patterns; (11) \textbf{Atten-Mixer}~\cite{Zhang@WSDM2023} leverages multi-level user intents to achieve multi-level reasoning on item co-occurrence patterns. 

\paratitle{LLM-based method:} (12) \textbf{P5}~\cite{P5} is a representative effort incorporating large language model (LLM) into recommendation tasks. It transforms user-item interactions into natural language sequences, pre-trains a Transformer architecture on these sequences, and uses adaptive prompts to provide personalized recommendations. 

\subsubsection{Evaluation ‌Protocol}

In this study, we adopt a full-ranking training paradigm across the entire item set. In this manner, all models are tasked with ranking all items based on their predicted probability of being interacted with next, given a specific user-item interaction sequence. The top-k items with the highest predicted interaction probability are then formulated as the recommendation list, denoted as $rec$ = [$y_1, y_2, ..., y_k$], where $y_i$ $\in$ $\mathcal{I}$.

Following the common practice~\cite{GRU4Rec,NARM,SR-GNN,BERT4Rec,Zhang@WSDM2023}, we evaluate the performance of all methods using two widely adopted metrics: \textbf{Prec@k} (Precision), which measures the proportion of cases where the label items appear within the recommendation lists; and \textbf{MRR@k} (Mean Reciprocal Rank), which calculates the average of the reciprocal ranks of the label items within the recommendation lists. As in~\cite{NARM,BERT4Rec,SR-GNN}, we set k = 10 in this work. For both metrics, higher values indicate better performance. 

\subsubsection{Implement Details}
To fairly evaluate the ability of current methods in capturing collaborative information, we adhere to the implementations described in their original papers. For the main hyper-parameters of these models, we optimize them using grid search on the validation set. Specifically, the embedding size for all models is selected from the set $\{64, 128, 256, 512, 1024\}$. To guarantee that these models focus solely on handling collaborative information, we use only the user-item interactions (\ie item ID sequences) as input, without exposing the models to any additional information, such as item text or images. For the LLM-based model P5, we implement it with base version where the encoder and decoder both have 12 Transformer blocks. The source codes are available online\footnote{\url{https://github.com/Zhang-xiaokun/CI_Investigation}}.

\subsection{Overall performance}\label{sec_overall}

The overall performance of all 12 methods is presented in Table~\ref{performance}. To intuitively illustrate the performance differences, we have added a superscript to each model's result to signify its ranking within the respective metric. From the results presented in Table~\ref{performance}, the following observations can be made:

First, recommendation methods exhibit significant performance differences across different datasets. For example, most neural methods achieve over 50\% accuracy on the Yoochoose dataset in terms of Prec@10, whereas their performance on the Cellphones does not exceed 20\%. Although this phenomenon is commonly observed, it remains underexplored in the existing literature. Interestingly, clues to this discrepancy can be gained by examining the distribution of CRs. Referring to~\fig~\ref{labelD} and Table~\ref{performance}, we can observe that \textbf{the performance of RS in a given dataset tends to be positively correlated with the proportion of simple CRs relevant to label items in the test set}. Specifically, compared to the Cellphones dataset, there are more 0-hop CRs relevant to label items in the test set of Yoochoose, \ie 67.04\% in Yoochoose versus 18.31\% in Cellphones. This observation is intuitive, as simple CRs are easier to detect and exploit, enabling models to perform better when such relations are more prevalent in the test data. In addition, it also underscores the significance of collaborative information in recommendation tasks and validates the effectiveness of our proposed quantitative definition for this crucial resource.

\begin{table}[t]
\small
\tabcolsep 0.02in 
\centering
\caption{Statistics of test sequences in which the label item holds only one CR with all items in the sequence.}
\begin{tabular}{c ccc ccc}
\toprule
Datasets        & Grocery & Cellphones & Cosmetics & Diginetica & Yoochoose & Tmall \\
\midrule
0-hop CR        & 1,256    &628    &1,222    &2,274    &14,621  &237   \\
1-hop CR        & 905    &1,217    &827    &1,015    &1,454    &205   \\
2-hop CR        & 543    &967    &77    &336    &90 &364     \\
others          & 42    &56    &4    &49    &6  &191 \\
\bottomrule
\end{tabular}

\label{purestatistics}
\end{table}

Second, advanced neural models, \eg BERT4Rec and MSGIFSR, generally outperform traditional methods. This is a trend commonly observed in previous studies~\cite{NARM, BERT4Rec}. By investigating collaborative information, we offer a new perspective to explain this phenomenon. Traditional methods are limited by simple item similarity measures, such as statistical item co-occurrence frequency. Thus, they can primarily handle basic CRs like 0-hop CR. However, as shown in~\fig~\ref{overallD}, \textbf{simple CRs only account for a small portion of item-item CRs, leading to the inability of traditional methods in copying with most items.} In contrast, neural models, with their powerful representation capabilities, can explore more complex CRs, enabling them to achieve superior prediction accuracy.

Third, MSGIFSR achieves exceptional performance across various benchmarks, outperforming all other methods in most cases. This approach extracts groups of consecutive items from a sequence to construct multiple sub-graphs, enabling it to capture fine-grained co-occurrence patterns within user-item interactions. By adopting this paradigm, MSGIFSR is able to explore various CRs among items, significantly boosting its recommendation performance. This highlights \textbf{the importance of handling diverse CRs in improving recommendation effectiveness}.

Finally, the LLM-based method, \ie P5, is outperformed by neural models in most cases. 
This suggests that, contrary to its success in various tasks of natural language processing, LLMs do not achieve anticipated breakthroughs in the recommendation task.
While possessing a wealth of world knowledge and excel at encoding modality information such as text or images, \textbf{LLMs struggle to capture collaborative knowledge from user-item interactions} (i.e., item ID sequences). This limitation leads to their under-performance in performing personalization.

\subsection{Impact of Diverse CRs on RS}\label{sec_impactofCR}

\begin{figure}[t]
  \centering
  \includegraphics[width=0.99\linewidth]{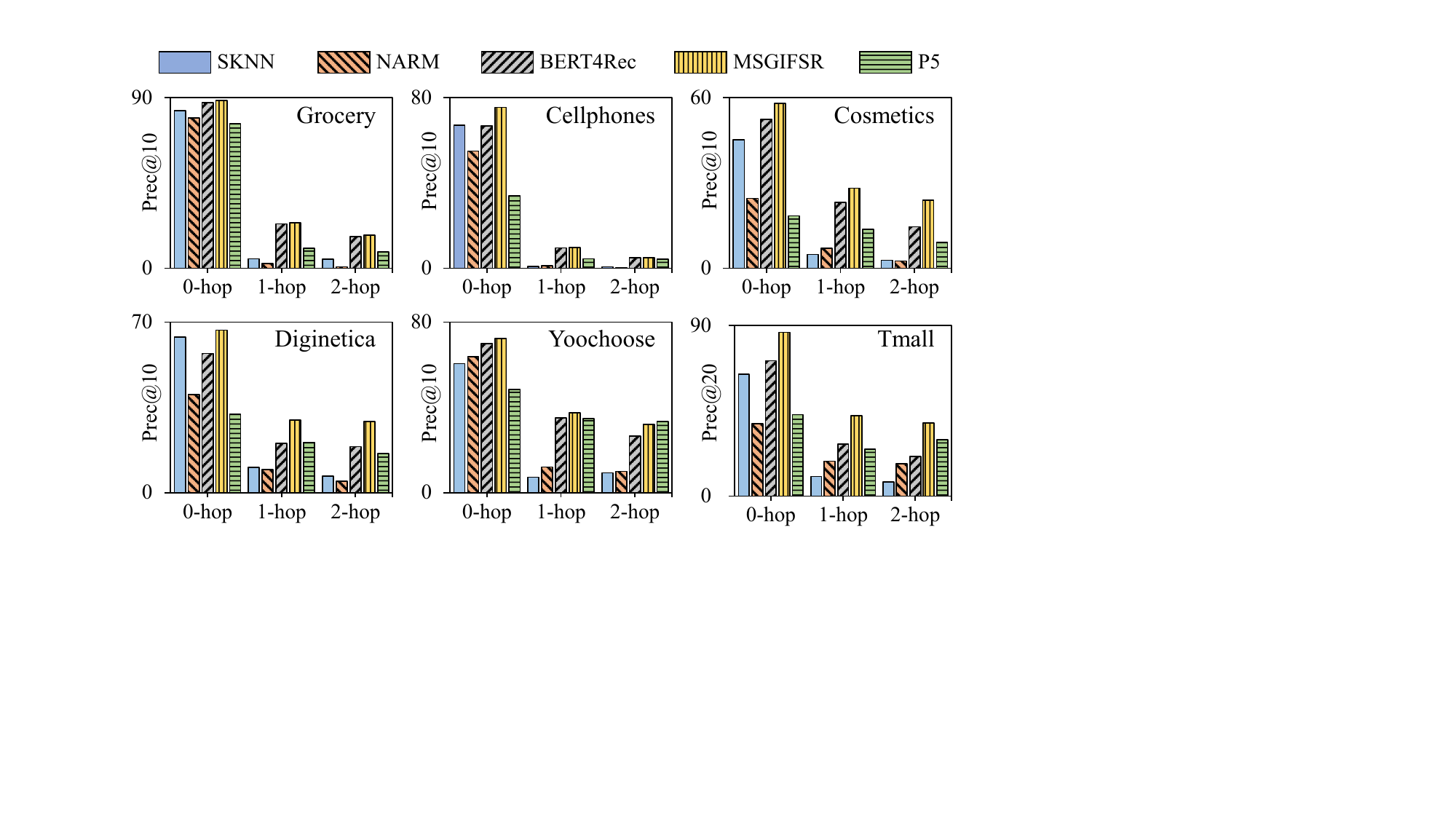}
  \caption{Impact of diverse CRs on RS.}\label{predictCR}
\end{figure}

In this section, we delve into the impact of diverse types of CRs on RS performance. As discussed in Section~\ref{CRdistribution}, a label item can possess several types of CRs with different items in a sequence, making it challenging to identify which specific type of CRs RS methods have captured when generating recommendations. To isolate the effect of a particular CR on RS performance, we focus on `pure' samples, where the label item has only one single type of CR about all items in the sequence. The number of pure samples for each CR type across various benchmarks is presented in Table~\ref{purestatistics}. The performance of several representative methods, including the traditional SKNN, the neural models NARM, BERT4Rec, as well as MSGIFSR, and the LLM-based method P5, under this pure setting is shown in~\fig~\ref{predictCR}. Note that, we omit results of the `others' case due to the limited number of samples in this situation, which could not provide statistically significant results. We can obtain following observations from~\fig~\ref{predictCR}:

First, RS methods present significant performance differences across different types of CRs, with notably better prediction accuracy on simpler CRs compared to more complex ones. In particular, their performance on 0-hop CR is much higher than that on other types of CR. This suggests that \textbf{RS methods excel at capturing simple CRs, but struggle with more complex ones.} This is intuitive, as the simple CR represents straightforward item co-occurrence patterns, which are easy to detect. In contrast, as discussed in Section~\ref{chaofCR}, complex CRs stem from simple CR and embody richer, subtler patterns, making them harder for RS to understand and effectively utilize.

Second, as previously discussed, neural models generally outperform traditional methods. Through a fine-grained analysis of model performance across different CR types, we gain further insights into this phenomenon. \textbf{The superiority of advanced methods over traditional techniques mainly stems from their ability in handling complex CRs.} For example, while NARM (neural model) is defeated by SKNN (traditional method) on 0-hop CR in Tmall dataset, it significantly outperforms SKNN on 1-and 2-hop CRs. We can attribute this to the sophisticated architectures of advanced models, like RNNs or GNNs, which enable them to capture complex CRs, thereby enhancing their overall performance.

Third, the performance gap of P5 across different CR types is the smallest among all methods, as shown in Cosmetics dataset. It indicates that \textbf{the LLM-based method demonstrates relatively stable performance across diverse CRs}. Unlike current methods, which rely solely on recommendation data to provide personalization, LLM-based approaches incorporate a wealth of world knowledge, acquired through pre-training on massive language data. This extra information might provide new perspectives for LLM to understand diverse CRs, leading to its stable performance across various CRs.  

\subsection{Direct v.s. Indirect CRs}\label{sec_direct}

\begin{table}[t]
\small
\tabcolsep 0.01in 
\centering
\caption{Statistics of test sequences under direct and indirect CR situations.}
\begin{tabular}{c ccc ccc}
\toprule
Datasets        & Grocery & Cellphones & Cosmetics & Diginetica & Yoochoose & Tmall \\
\midrule
direct        & 2,061    &1,437    &3,385    &4,821    &17,645  &796	\\
indirect      & 1,994      &3,173    &1,212    &1,925    &1,742    &1,273 \\
\bottomrule
\end{tabular}

\label{directSta}
\end{table}

\begin{figure}[t]
  \centering
  \includegraphics[width=0.98\linewidth]{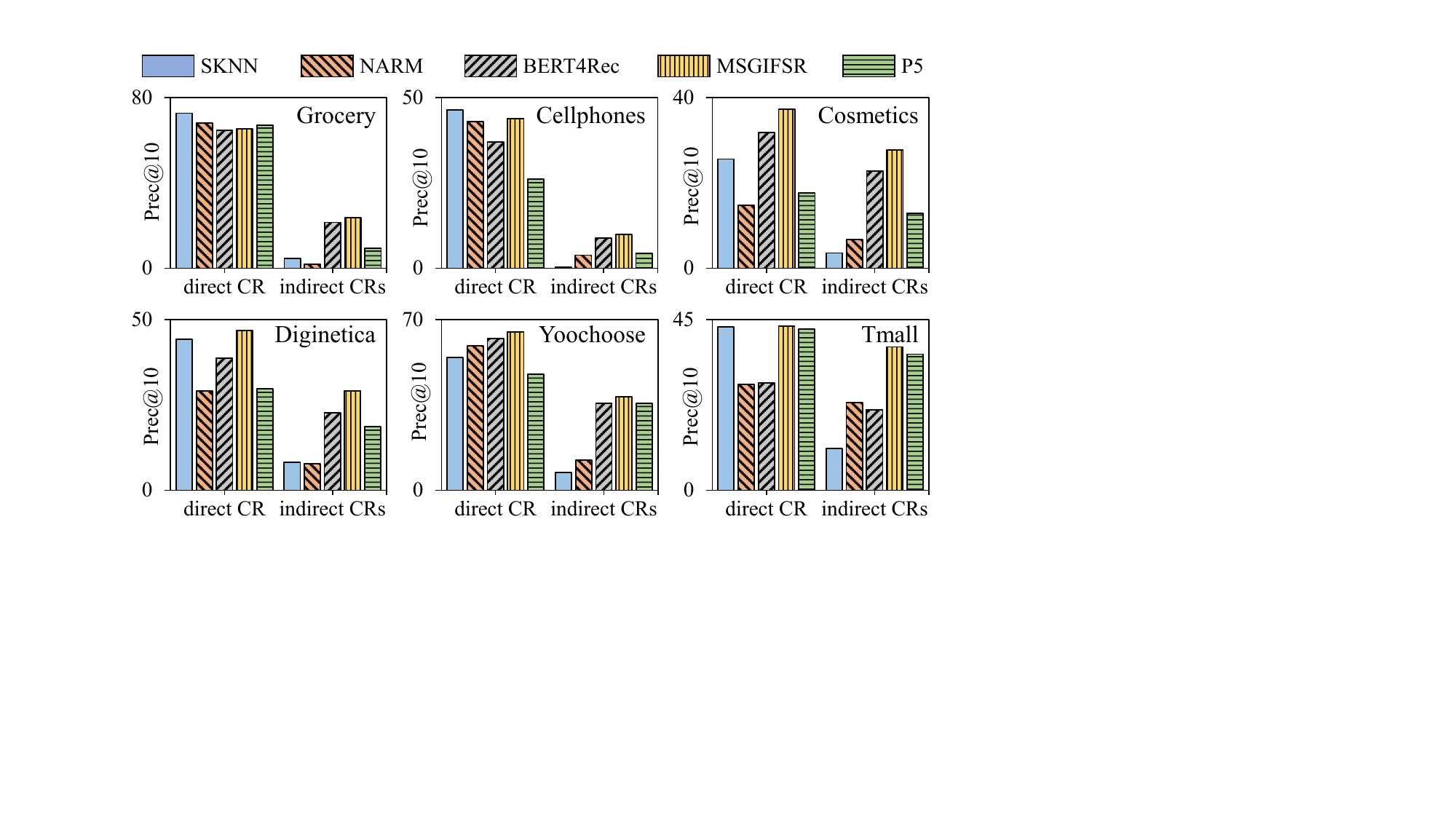}
  \caption{Impact of direct and indirect CRs on RS.}\label{direct}
\end{figure}

In the previous section, we find that RS achieve superior results on simple CRs over complex ones, particularly excelling at the 0-hop CR (direct CR). Thus, we further investigate the impact of direct and indirect CRs on RS performance. Specifically, we divide the test sequences of a benchmark into two groups: (1) direct CR, where the label item has a 0-hop CR with at least one item in the sequence; and (2) indirect CRs, where the label item does NOT hold the 0-hop CR with any items in the sequence. These two groups are mutually exclusive and collectively form the entire test set. The statistics of these groups are presented in Table~\ref{directSta}.

From~\fig~\ref{direct}, we can observe that RS perform significantly better on direct CR than on indirect CRs. This suggests that \textbf{the strength of current recommendation methods lies primarily in their ability to capture and leverage direct CR between items.} Unfortunately, their performance in capturing indirect CRs remains limited. This highlights an important insight for the research community: \textbf{improving RS ability to handle complex, indirect CRs could significantly enhance its overall recommendation performance.}

\subsection{Proportion of Different CRs in Predictions}\label{sec_predictCR}

\begin{figure}[t]
  \centering
  \includegraphics[width=0.99\linewidth]{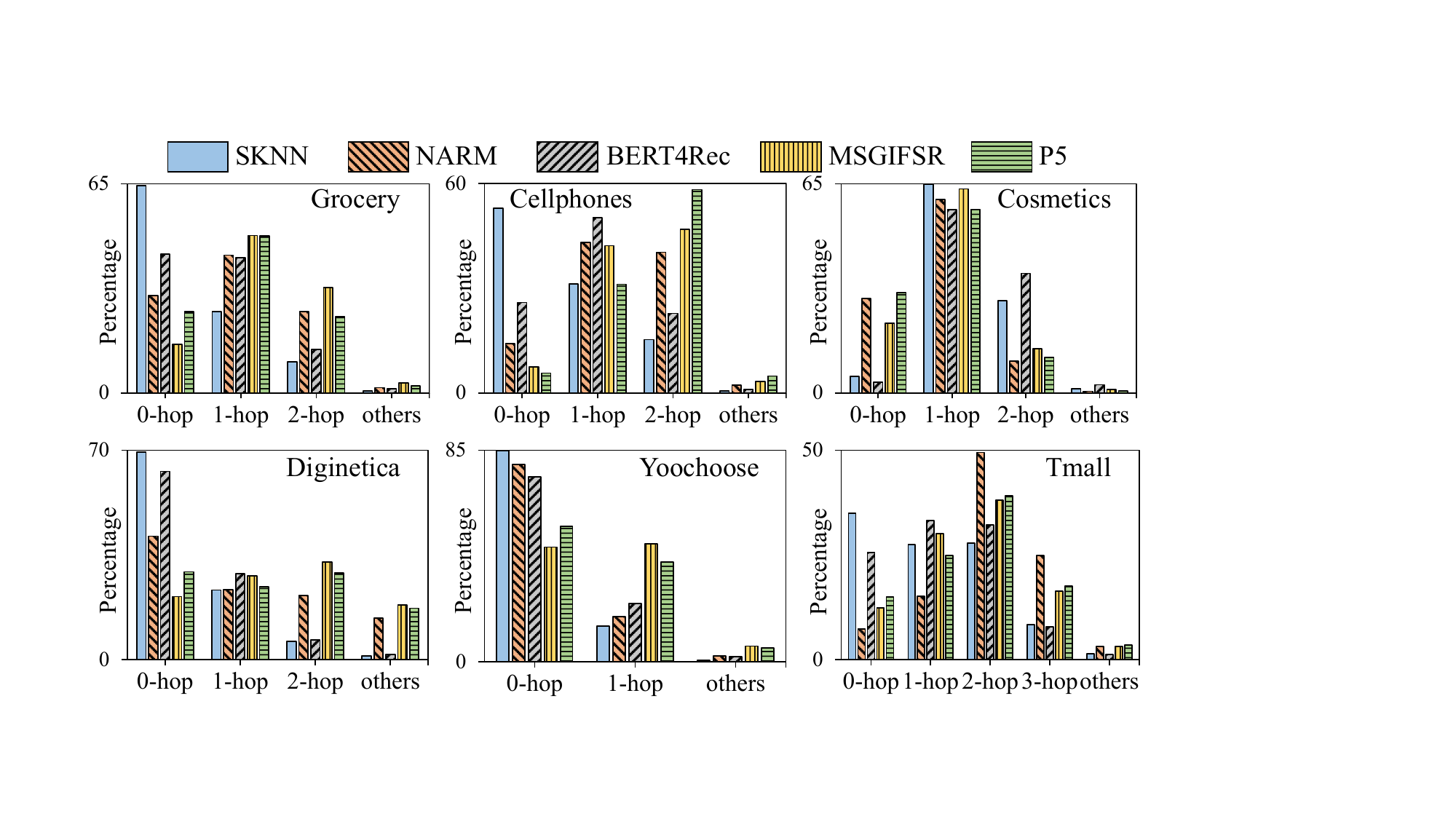}
  \caption{Proportion of different CRs in predictions.}\label{cr_pro}
\end{figure}

This section aims to examine the ability of RS in utilizing various CRs. Specifically, we analyze which types of CRs recommendation models tend to rely on when making predictions. Given a user-item interaction sequence $\mathcal{S}$ = [$x_1, x_2, \cdots, x_m$] in a test set, a model generates a recommendation list containing k items, \ie $rec$ = [$y_1, y_2, \cdots, y_k$], where k=10. We record the CR between each predicted item $y_i \in rec$ and each item $x_i \in \mathcal{S}$, where this CR serves as the clue the model relies on to make the prediction. We then analyze these CRs corresponding to all recommendation lists within the test set, and compute the proportion of each type CR. As shown in~\fig~\ref{cr_pro}, following observations are noted.


First, we have considered the scenario where the predicted item has no CRs with any items in $\mathcal{S}$. However, this scenario does not appear in any of the datasets, so it is not included in the figure. In fact, \textbf{all predicted items are found to have certain CRs with items that the user has interacted with}. This observation indicates that \textbf{existing RS methods largely depend on CRs to deliver personalized services}, highlighting the crucial role of CRs in this task. It also validates the effectiveness of our proposed quantitative definition of collaborative information.

Second, the proportion of simple CRs (such as 0-hop and 1-hop CRs) is generally much higher than that of complex CRs (\eg `Others') in the predictions made by all methods. This indicates that \textbf{current RS methods predominantly rely on simple CRs when generating recommendations.} This is intuitive, as simple CRs are easier to detect and leverage compared to complex ones. Additionally, this reinforces the limitations of current methods in effectively utilizing complex CRs. 

Third, the proportion of simple CRs in results generated by traditional methods is much higher than in those produced by advanced models, while the opposite is true for complex CRs. This signifies that, \textbf{with the advancement of techniques, \eg from nearest-neighbor methods to neural architectures, the ability of RS to utilize complex CRs is improving.} Surprisingly, the results generated by LLMs show a higher proportion of complex CRs compared to other methods, as exemplified in the Cellphones dataset. \textbf{This underscores the potential of LLMs in understanding and leveraging complex CRs for recommendations}, a capability not previously explored in existing literature.

\section{Challenges and Future Directions}

Building on the above observations of collaborative information, this section further highlights remaining challenges and points out potential directions for future research.

\paratitle{C1: How to address the imbalanced CR distribution?}

The CR distribution exhibits significant imbalances in two aspects: inter-CR and intra-CR. For the inter-CR, the number of item pairs included in each type of CR varies greatly, as illustrated in~\fig~\ref{overallD}. As to the intra-CR, the co-occurrence frequency of item pairs within a certain type of CR follows a long-tail pattern, as plotted in~\fig~\ref{co_0}. These highly skewed distributions pose a challenge for RS, making it difficult to effectively capture the full range of CRs from user-item interactions. This issue is particularly deteriorated in neural network-based methods, where imbalanced data can negatively affect both predictive accuracy and model scalability. Therefore, addressing the imbalanced CR distribution could emerge as an important direction for future research. 

\paratitle{C2: How to enhance RS ability in capturing complex CRs?}

The ability of existing RS to handle diverse types of CRs varies significantly. As highlighted in Section~\ref{sec_impactofCR} and~\ref{sec_direct}, these methods generally perform much better on simple CRs than on complex ones. Additionally, as discussed in Section~\ref{sec_predictCR}, current methods predominantly tend to rely on simple CRs for making predictions, rather than leveraging complex CRs. These findings underscore the challenges associated with effectively utilizing complex CRs. In the meantime, this also suggests that improving RS ability to handle such complex relations could lead to a substantial enhancement in overall performance~\cite{Harald@RecSys2021}. As a result, enhancing RS capability to process complex CRs presents a promising direction for RS.

\paratitle{C3: How to inject collaborative knowledge into LLMs?}

LLMs are pre-trained on vast corpora of text data, endowing them with extensive world knowledge~\cite{Sun@SIGIR2024,Kim@KDD2024,Wang@CIKM2024}. However, they often struggle to capture the collaborative knowledge inherent in user-item interactions, which is crucial for personalized recommendations~\cite{Hu@WWW2024, Zheng@ICDE2024, Yang@SIGIR2024}. As discussed in Section~\ref{sec_overall}, this limitation leads to their suboptimal performance in the task. Our investigation into collaborative information has deepened our understanding of this crucial resource. These insights open up new opportunities to inject collaborative knowledge into LLMs. By effectively combining the strengths of LLMs in processing language with the power of collaborative information, future RS can deliver more accurate, personalized, and even explainable recommendations.

\section{Related Work}


\paratitle{Traditional techniques.} Nearest-neighbor-based approaches aim to identify similar users or items to recommend based on their proximity to a given user or item, like Item-KNN~\cite{IKNN} and SKNN approaches~\cite{Jannach@RecSys2017, Garg@SIGIR2019}. Matrix factorization~\cite{SalakhutdinovNIPS2007} learns user and item vectors from the user-item interaction matrix, formulating recommendations based on the similarity between these vectors. Markov chain-based methods, like Markov Decision Process (MDP)~\cite{MDP}, frame the recommendation problem as a sequential optimization task, modeling the user-item interaction sequences via an MDP.


\paratitle{Neural networks.} Due to their powerful representation capabilities, various neural architectures are employed to enhance recommendation systems~\cite{Cheng@WWW2022}. Early efforts are dedicated to applying Recurrent Neural Network (RNN) and its variants to model user-item interaction sequences due to its inherent capability in handling sequential data, \eg GRU4Rec~\cite{GRU4Rec}. Attention mechanism is another popular choice in RS, helping to distinguish the importance of different items within user-item interactions, as demonstrated by SASRec~\cite{SASRec} and BERT4Rec~\cite{BERT4Rec}. Moreover, some efforts build user-item interaction graphs and capture item transition patterns using Graph Neural Networks~\cite{SR-GNN,LESSR,DHCN}. Some methods further enhance GNNs by dividing user-item interaction sequences into multiple subsequences and constructing sub-graphs for each sequence to capture item co-occurrence patterns more effectively, such as MSGIFSR~\cite{guo@WSDM2022} and Atten-Mixer~\cite{Zhang@WSDM2023}.

\paratitle{LLM-based methods.} Recently, there is a growing trend to incorporate LLMs into recommendation systems~\cite{Yang@SIGIR2024,Sun@SIGIR2024,Zhao@SIGIR2024}. However, LLMs face challenges in encapsulating collaborative knowledge from user-item interactions, as they are not trained on specialized recommendation data, often leading to their suboptimal performance in this domain~\cite{Liu@CIKM2024}. To address this limitation, recent efforts typically process user-item interactions through established neural models to extract collaborative knowledge, while also leveraging LLMs to handle additional information, \eg item text or images, in order to derive semantic insights. Afterwards, these two types of knowledge are combined to generate personalized recommendations~\cite{Zheng@ICDE2024,Zhu@WWW2024,Kim@KDD2024,Wang@CIKM2024}.

\section{Conclusions}

Collaborative information has dominated the study of recommender systems over several decades. However, there has been a lack of comprehensive research on this crucial resource within the community. To address this gap, this work is dedicated to systematically investigating collaborative information from several key aspects: (1) we clarify collaborative information in terms of item co-occurrence patterns, identify its main characteristics including Transitivity, Hierarchy as well as Redundancy, and give a quantitative definition of this concept; (2) we explore the manifestation of diverse collaborative relations within user-item interactions, examining their distributions from various aspects; and (3) we evaluate the impact of diverse collaborative relations on the performance of various recommendation algorithms. Through this systematic investigation, we uncover many novel insights into the collaborative information within recommendation scenarios. After delving into these findings, we further discuss the remaining challenges and present potential directions for future research.

\section*{Acknowledgement}
This work is supported by the Early Career Scheme (No.CityU 21219323) and the General Research Fund (No.CityU 11220324) of the University Grants Committee (UGC), the NSFC Young Scientists Fund (No.9240127), and the Donation for Research Projects (No.9229164 and No.9220187). This work is also supported by the Key Laboratory of Social Computing and Cognitive Intelligence (Dalian University of Technology), Ministry of Education (SCCI2025YB04).


\newpage

\bibliographystyle{ACM-Reference-Format}
\bibliography{ref}

\newpage

\end{document}